\documentclass[twocolumn]{aastex63}
\usepackage{graphics,epsf}
\usepackage{amsmath}                
\usepackage{amsfonts}               
\usepackage{amssymb}                
\usepackage{epsfig}                 
\usepackage{appendix}
\usepackage{graphicx}
\usepackage{float}
\usepackage{color}
\usepackage{multirow}
\usepackage{colortbl}
\usepackage[para,online,flushleft]{threeparttable}
\usepackage{url}

\hypersetup{citecolor=blue, 
            linkcolor=red, 
            menucolor=blue, 
            urlcolor=blue}  
            
\usepackage{hyperref}

\newcommand{\cm}{{~\rm cm}}
\newcommand{\m}{{~\rm m}}
\newcommand{\km}{{~\rm km}}
\newcommand{\s}{{~\rm s}}
\newcommand{\h}{{~\rm h}}

\newcommand{\g}{{~\rm g}}
\newcommand{\G}{{~\rm G}}
\newcommand{\K}{{~\rm K}}
\newcommand{\erg}{{~\rm erg}}
\newcommand{\yr}{{~\rm yr}}

\newcommand{\pc}{{~\rm pc}}

\newcommand{\AU}{{~\rm AU}}

\newcommand{\days}{{~\rm days}}

\begin{document}

\title{The core degenerate scenario for the type Ia supernova SN 2020eyj}


\author{Noam Soker}
\affiliation{Department of Physics, Technion, Haifa, 3200003, Israel; soker@physics.technion.ac.il; ealeal44@technion.ac.il}

\author{Ealeal Bear}
\affiliation{Department of Physics, Technion, Haifa, 3200003, Israel; soker@physics.technion.ac.il; ealeal44@technion.ac.il}

\begin{abstract}
We argue that the core degenerate (CD) scenario of type Ia supernovae (SNe Ia) can explain the compact helium-rich circumstellar material (CSM) of SN 2020eyj. In the new channel of the CD scenario that we propose there are two major common envelope evolution (CEE) phases. After the white dwarf (WD) companion removes the hydrogen-rich envelope of the asymptotic giant branch star its spiralling-in halts at few solar radii from the core, rather than continuing to the carbon-oxygen (CO) core as in the hydrogen-rich SNe Ia-CSM CD scenario. Only hundreds to tens of thousands of years later, after the helium-rich core expands, does the WD enters a CEE with the helium-rich layer. By that time the hydrogen-rich envelope is at a large distance from the center. The WD merges with the CO core during the second CEE phase, and only after a merger to explosion delay (MED) time of weeks to tens of years the merger remnant explodes. The SN Ia ejecta collides with a helium-rich CSM at tens to hundreds of AU. We follow the evolution of two stellar models with initial masses of $5M_\odot$ and $7M_\odot$ to their asymptotic giant branch phase when they are supposed to engulf the WD companion.  We find that there is a sufficiently massive CO core to merge with the WD in the frame of the CD scenario as well as a massive helium-rich layer, $\simeq 0.3-1M_\odot$, to account for the helium-rich CSM of SN 2020eyj. 
\end{abstract}


\section{INTRODUCTION}
\label{sec:intro}

There are several theoretical type Ia supernova (SN Ia) scenarios (e.g., \citealt{Hoeflich2017, LivioMazzali2018, Soker2018Rev, Wang2018,  Jhaetal2019NatAs, RuizLapuente2019, Soker2019Rev, Ruiter2020} for some recent reviews). \cite{Soker2019Rev} classify them as follows (we consider only scenarios that involve binary interaction). 
\begin{enumerate}
\item The \textit{core-degenerate (CD) scenario} that involves a common envelope evolution (CEE) with the merger of a CO WD or a HeCO WD companion with the core of a massive asymptotic giant branch (AGB) star. The (close to) Chandrasekhar-mass WD remnant explodes later (e.g., \citealt{KashiSoker2011, Ilkov2013, AznarSiguanetal2015}). 
\item The \textit{double degenerate (DD) scenarios and the DD-MED scenario} where two WDs merge (e.g., \citealt{Webbink1984, Iben1984}) after losing energy to gravitational waves. There are different channels of the DD scenario. For example, the explosion itself might take place during the merger process (e.g., \citealt{Pakmoretal2011, Liuetal2016, Ablimitetal2016}), including the ignition of helium first (e.g., \citealt{YungelsonKuranov2017, Zenatietal2019, Peretsetal2019}), or at a later time (merger to explosion delay: MED) time (e.g., \citealt{LorenAguilar2009, vanKerkwijk2010, Pakmor2013, Levanonetal2015, LevanonSoker2019, Neopaneetal2022}). 
\item The \textit{double-detonation (DDet) scenario} attributes the explosion of a CO WD to the ignition of helium that the CO WD accretes (e.g., \citealt{Woosley1994, Livne1995, Shenetal2018}). It has several channels, including of peculiar SNe Ia like the core merger detonation (CMD) scenario \citep{Ablimit2021}. 
\item The \textit{single degenerate (SD) scenario} involves a CO WD that accretes hydrogen-rich gas from a non-degenerate donor and explodes as its mass is about the Chandrasekhar mass limit (e.g., \citealt{Whelan1973, HanPodsiadlowski2004, Orio2006, Wangetal2009, MengPodsiadlowski2018}).
This scenario has several channels. Explosion might occur as soon as the CO WD reaches close to the Chandrasekhar mass limit or the explosion occurs with a long time delay after the CO WD losses angular momentum (e.g., \citealt{Piersantietal2003, DiStefanoetal2011, Justham2011}). The accretion process might vary, like being from a main sequence donor, from a giant donor, or inside a common envelope in the common-envelope wind model (e.g., \citealt{Cuietal2022}).  
\item  The \textit{WD-WD collision (WWC) scenario} attributes the explosion ignition to the collision of two WDs (e.g., \citealt{Raskinetal2009, Rosswogetal2009, Kushniretal2013, AznarSiguanetal2014}). The WWC scenario contributes  $<1 \%$ of all normal (non-peculiar) SNe Ia (e.g., \citealt{Toonenetal2018, HallakounMaoz2019, HamersThompson2019, GrishinPerets2022}).    
\end{enumerate}

Different recent papers consider different SN Ia scenarios from the above list to be the dominant scenario(s) 
(e.g., some papers from last two year, \citealt{CuiXetal2020, Zouetal2020, Blondinetal2021, Clarketal2021,  Chandraetal2021, Hakobyanetal2021, Liuetal2021, MengLuo2021, Michaely2021, WangQetal2021, Yamaguchietal2021, Zengetal2021, Ablimit2022, Acharovaetal2022, AlanBilir2022, Barkhudaryan2022, Chanlaridisetal2022, Chuetal2022, CuiLi2022, Dimitriadisetal2022, Ferrandetal2022, Igoshevetal2022, Kosakowskietal2022, Kwoketal2022, Lachetal2022, Liuetal2022, LivnehKaatz2022, Mazzalietal2022, Pakmoretal2022, Patraetal2022, Piersantietal2022, RauPan2022, RuizLapuenteetal2022, Sanoetal2022, SharonKushnir2022, Shingles2022, Tiwarietal2022, DerKacyetal2023a, DerKacyetal2023b, IwataMaeda2023, Maedaetal2023, SwarubaRajamuthukumaretal2023}). 
This long list of papers that use different scenarios  implies that in analysing observations we should be aware of all scenarios, and not of only some of them. 

In a recent study \cite{Kooletal2022} describe their observations and interpretation of SN 2020eyj, a SN Ia that interacts with a helium-rich circumstellar material (CSM). They argue that their observations support the SD scenario with a helium non-degenerate donor star as the progenitor of SN 2020eyj. \cite{Kooletal2022} conclude that SN 2020eyj did not result from the DD scenario and therefore their observation ``leave the SD scenario as the only viable alternative.'' As we mentioned above, one must consider other scenarios as well when analysing SNe Ia. We do just that in this paper by considering the CD scenario.  

In section \ref{sec:SN2020eyj} we review relevant properties of SN 2020eyj and argue that the SD scenario cannot explain some of these properties. In section \ref{sec:TheCDscenario} we propose instead that the CD scenario where the WD enters a CEE with an expanding helium-rich envelope at a late time best explains the properties of SN 2020eyj. 
We summarize this study in section \ref{sec:Summary}. 

\section{Relevant properties of SN 2020eyj}
\label{sec:SN2020eyj}

\cite{Kooletal2022} describe in details the properties of SN 2020eyj. We describe the relevant ones for our study of the CD scenario. 
 
In the first $\simeq 50 \days$ post-discovery SN 2020eyj behaves as a normal SN Ia, although with ejecta velocities on the slow side of SNe Ia and peak luminosity on the fainter end of a sample of SN Ia that interact with a circumstellar material (SNe Ia-CSM; \citealt{Silvermanetal2013}). 

Then, from day $\simeq +50$ to day $\simeq +250$ post-discovery the lightcurve flatten, resuming decline at 
day $\simeq +250$. \cite{Kooletal2022} attribute this lightcurve plateau to the interaction of the ejecta with a CSM. The spectrum shows this CSM to be helium-rich and hydrogen-poor, and the start time of the interaction suggests an inner CSM radius of $r_{\rm CSM,in}\simeq 4 \times 10^{15} \cm$. \cite{Kooletal2022} calculate the CSM mass for a shell model and for a wind, and deduce CSM masses for the two cases of $M_{\rm csm}({\rm shell)} = 0.36 M_\odot$ and $M_{\rm csm}({\rm wind}) = 0.3-1 M_\odot$, respectively. 

\cite{Kooletal2022} find that SN 2020eyj shows strong similarities to the SN Ia-CSM PTF11kx, including the CSM mass. \cite{Sokeretal2013} argue that the SD scenario cannot account for such a mass in the case of PTF11kx. Definitely nova eruptions cannot account for such a large mass. The case for a wind from the progenitor is more problematic in the case of SN 2020eyj than for PTF11kx because the donor is a helium-star and therefore cannot be as large as hydrogen-rich AGB stars. The progenitor of PTF11kx was hydrogen-rich which allows in principle for a giant donor. \cite{Kooletal2022} note this difficulty and speculate that the He-rich CSM of SN 2020eyj is in a long-lived disk, as in the only known helium nova (He-rich but H-free nova ejecta) V445~Puppis (e.g., \citealt{Nyamaietal2021}). The dust mass in V445~Puppis, however, seems to be only $\approx 5 \times 10^{-4} M_\odot$ \citep{Shimamotoetal2017}, while \cite{Kooletal2022} estimate the dust mass in SN 2020eyj to be $\approx 0.01 M_\odot$.

The concentration of the CSM in a long-lived disk raises other problems. The first problem is the ratio of the CSM radius to the orbital separation of the binary system. The radius of a helium star donor is not large, $\la 1 \AU$. Therefore, for a large mass transfer rate the orbital separation must be $a \la 3 \AU$, depending on eccentricity. The orbital period of V445~Puppis of $0.65 \days$ \citep{Goranskijetal2010} suggests an orbital separation of only ${\rm several}\times R_\odot$.   
The ratio of the CSM inner radius to the binary orbital separation is then $r_{\rm CSM,in}/a \approx100- 10^4$. It is not clear how such a binary can supply the angular momentum to support such a long-lived massive disk, $M_{\rm csm}({\rm shell)} \ga 0.3 M_\odot$. 

Another problem is that only a small fraction of the ejecta interacts with a CSM that is concentrated in a long-lived disk. For example, the late spectra of SN 2020eyj lack the O\textsc{i} 7774\AA{} 
emission line, as in other SNe-CSM. If only a small fraction of the ejecta interacts with the CSM, we might expect the rest of the ejecta to have a spectrum as non-interacting SNe Ia. It is still possible that the He-rich CSM is in a bipolar structure, e.g., a very thick equatorial outflow with lower-density along the poles.

There are indirect arguments against the SD scenario with a He-rich donor, the one that \cite{Kooletal2022} prefers. One is that such an explosion leaves behind a bright remnant, the He-rich donor. No such companion exists in CCSNRs in our Galaxy.  Another argument is that to have the huge mass transfer rate that \cite{Kooletal2022} consider, $\ga 10^{-3} M_\odot \yr^{-1}$, the orbital separation cannot be much larger than the radius of the donor. This in turn implies that a large fraction of the ejecta interacts with the He-rich envelope of the donor and entrain some He-rich gas and accelerates the helium to high velocities. Such entrained helium is not observed as there are no helium lines at early times.

\section{A core degenerate (CD) scenario for SN 2020eyj}
\label{sec:TheCDscenario}
\subsection{The He-rich CSM CD scenario}
\label{subsec:HeRichScenario}
There is a large group of SNe Ia with CSM at explosion. Since there is a hot WD that ionizes the CSM the object is (or was) a planetary nebula. This group of SNe Ia is termed SNIPs, for SNe Ia inside planetary nebulae. SNe Ia-CSM compose a small (e.g., \citealt{Soker2022Delay}) subgroup of SNIPs where at least part of the CSM is close enough to induce an early ejecta-CSM interaction. SN 2020eyj is a SNe Ia-CSM.
We here follow earlier claims (e.g., \citealt{Soker2022Delay} and references therein) that most, or even all, SNIPs, including SNIa-CSM, are formed according to the CD scenario, e.g., \cite{Sokeretal2013} for PTF11kx. The main argument is that the massive and compact CSM requires its ejection during a CEE.  However, the He-rich CSM of SN 2020eyj implies some differences in the CD scenario. 

We suggest that a CO WD (or a HeCO WD) spirals-in inside the hydrogen-rich giant envelope ejects that envelope, and ends the first CEE phase at a close orbit to the core of $a_{\rm f1} \simeq 5-10 R_\odot$. For that, according to the CD scenario, the AGB star should be massive at the onset of the CEE, $M_{\rm AGB} \ga 4 M_\odot$ (e.g. \citealt{IlkovSoker2012}).  At a later time $\Delta t_{\rm 2,CEE}$ a second CEE takes place. This occurs when evolution of the core and/or further tidal interaction force the CO WD to merge with the core. Namely, the CO WD enters a CEE inside the helium-rich outer parts of the core, ejects that helium-rich envelope, and merges with the CO core.  Based  on the evolution time between rapid envelope expansion, when the first CEE takes place, and until the helium core expands (section \ref{subsec:Core}), we  estimate that $\Delta t_{\rm 2,CEE} \simeq 10^3 - 10^5 \yr$.  

Some weeks to years later explosion takes place, as with other SNe Ia-CSM in the CD scenario.

 The short time delay from merger and envelope ejection to explosion might seem as a fine-tuning of the model. However, \cite{Soker2022Delay} estimates a flat distribution with time of explosions following merger during a CEE. This implies that many SNe Ia take place shortly after CEE. The recent observations of \cite{Sharmaetal2023} who estimate that $\simeq 0.02\%-0.2\%$ of SNe Ia interact with a compact CSM, support this estimate. A very recent example is SN 2020uem that \cite{Unoetal2023} study and suggest to be a SN Ia inside a massive and dense CSM in the frame of the CD scenario. Therefore, the scenario we propose is very rare, but it is not fine-tuned.  
 
In the present exploratory study we limit ourselves to show that cores with the appropriate properties, i.e., a massive CO core with a massive He layer, exist.  In a future study we will present the full evolutionary route. 

\subsection{The properties of the core}
\label{subsec:Core}

We evolve two stellar models with zero age main sequence (ZAMS) masses of $M_{\rm ZAMS}=5 M_\odot$ and $M_{\rm ZAMS}=7 M_\odot$, both with an initial metalicity of $Z=0.001$, with version 22.05.1 of the stellar evolution code Modules for Experiments in Stellar Astrophysics (\textsc{mesa}; \citealt{Paxtonetal2011, Paxtonetal2013, Paxtonetal2015, Paxtonetal2018, Paxtonetal2019}) in its single star mode.
\footnote{
The default capabilities  of \textsc{mesa}-single relay on the MESA EOS that is a blend of the OPAL \citep{RogersNayfonov2002}, SCVH
\citep{Saumonetal1995}, FreeEOS \citep{Irwin2004}, HELM \citep{TimmesSwesty2000}, PC \citep{PotekhinChabrier2010}, and Skye \citep{Jermynetal2021} EOSes. Radiative opacities are primarily from OPAL \citep{IglesiasRogers1993, IglesiasRogers1996}, with low-temperature data from \citet{Fergusonetal2005} and the high-temperature, Compton-scattering dominated regime by
\citet{Poutanen2017}.  Electron conduction opacities are from
\citet{Cassisietal2007} and \citet{Blouinetal2020}.
Nuclear reaction rates are from JINA REACLIB citep{Cyburtetal2010}, NACRE \citep{Anguloetal1999} and additional tabulated weak reaction rates \citealt{Fulleretal1985, Odaetal1994, Langankeetal2000}.  Screening is included via the prescription of \citet{Chugunovetal2007}. Thermal neutrino loss rates are from \citealt{Itohetal1996}.}

Changing the initial metalicity to $Z=0.02$ makes the stellar radius somewhat larger (by $\simeq 10 \%$) during the relevant evolutionary phases (see below), the CO core mass and the helium-rich layer mass in the core somewhat smaller (by $\simeq 25 \%$), and the expansion of the helium-rich core somewhat smaller (by $\simeq 35\%$ for the $5M_\odot$ model and by $\simeq 25\%$ for the $7M_\odot$ model). When we strip the hydrogen-rich envelope of a $7M_\odot$ model with $Z=0.02$ instead of $Z=0.001$ that we present in section \ref{subsec:CoreStripped}, we find that the helium-rich core suffers an even somewhat larger expansion in the second expansion phase to engulf the WD companion. We conclude that  changing the metalicity from  $Z=0.001$ to $Z=0.02$ changes nothing substantial in the evolutionary channel we study here. We therefore present the results only for the $Z=0.001$ cases.   

\begin{figure}[t]
	\centering
\includegraphics[trim=3.2cm 8.5cm 0.0cm 8.5cm ,clip, scale=0.60]{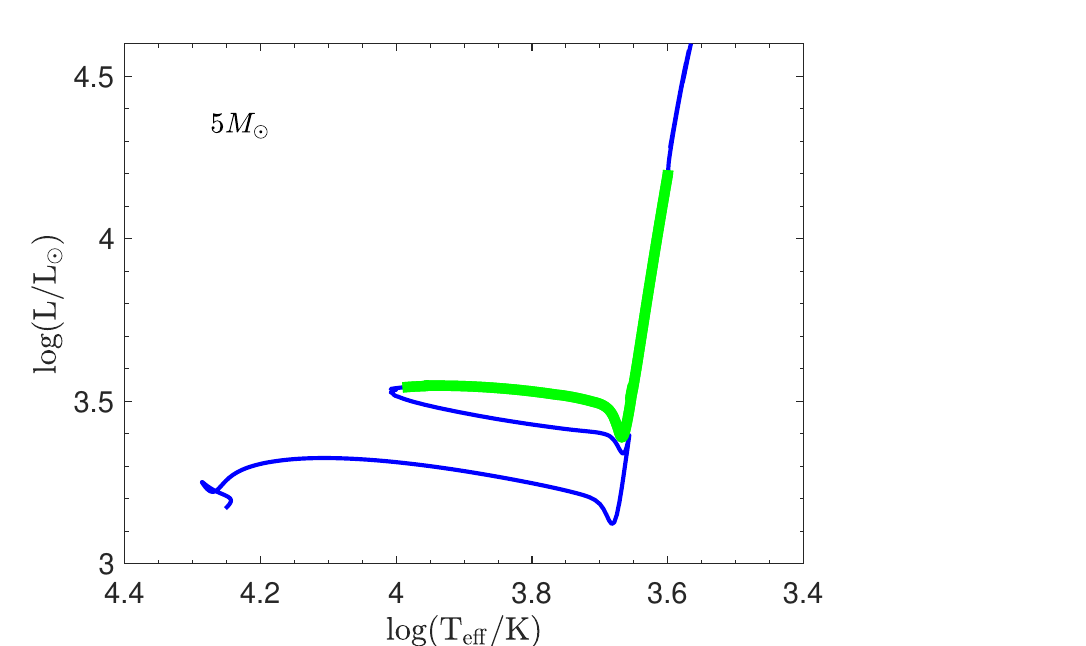}
 \\ 
\includegraphics[trim=3.2cm 8.5cm 0.0cm 8.5cm ,clip, scale=0.60]{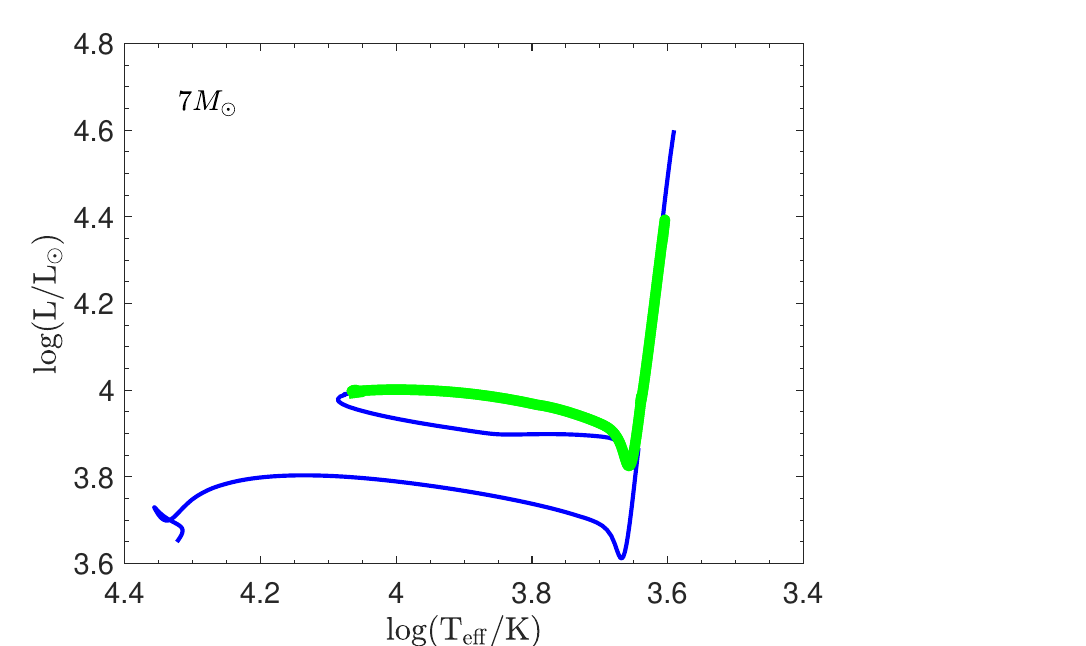}
\caption{The HR diagram from the ZAMS to the depletion of the massive helium-rich layer in the core of two stellar models with ZAMS masses of $M_{\rm ZAMS}=5 M_\odot$ (upper panel) and $M_{\rm ZAMS}=7 M_\odot$ (lower panel). The thick-green segment in each panel marks the evolutionary phase when the inner CO-core mass is $M_{\rm CO}>0.3 M_\odot$ and the helium-rich layer has a mass of $M_{\rm He-r} > 0.4 M_\odot$. }
\label{fig:HRdiagram}
\end{figure}

In Fig. \ref{fig:HRdiagram} we present the evolution of the two models on the HR diagram from their ZAMS to the time when the mass of helium-rich layer of the core becomes very small. 

We demand from the core of the star to have a large enough CO core so that a merger with the CO WD companion might set a SN Ia  explosion in the frame of the CD scenario. The CO WD mass is $M_{\rm WD} < 1.1 M_\odot$, and so we demand that the mass of the inner CO core be $M_{\rm CO}>0.3 M_\odot$. There is a possibility that the explosion will be of a sub-Chandrasekhar mass WD when helium ignites first. For example, the less-dense object in the binary, which is either the core or the WD, has a helium layer. The denser component of the binary tidally destroys that object, and the He-carbon mixture around the denser component is ignited first and triggers an explosion in the CO denser component (e.g., \citealt{Peretsetal2019}; but see \citealt{Pakmoretal2021} for another possible outcome). In both cases of the CD scenario, the Chandrasekhar mass explosion or the sub-Chandrasekhar mass explosion, we require some delay time to allow the ejected helium to form the CSM at $r> r_{\rm CSM,in}\simeq 4 \times 10^{15} \cm$. For the specific case of SN 2020eyj we require that the He-rich layer in the core, which the merger process ejects, be as massive as the CSM that \cite{Kooletal2022} infer, i.e., $M_{\rm He-r} \ga M_{\rm csm} > 0.3 M_\odot$ (section \ref{sec:SN2020eyj}). We demand here $M_{\rm He-r} > 0.4 M_\odot$. We mark by a thick-green line on the two panels of Fig. \ref{fig:HRdiagram} the evolutionary phase when the core obeys both conditions. 
 
\begin{figure}[]
	\centering
\includegraphics[trim=3.7cm 8.5cm 0.0cm 8.5cm ,clip, scale=0.59]{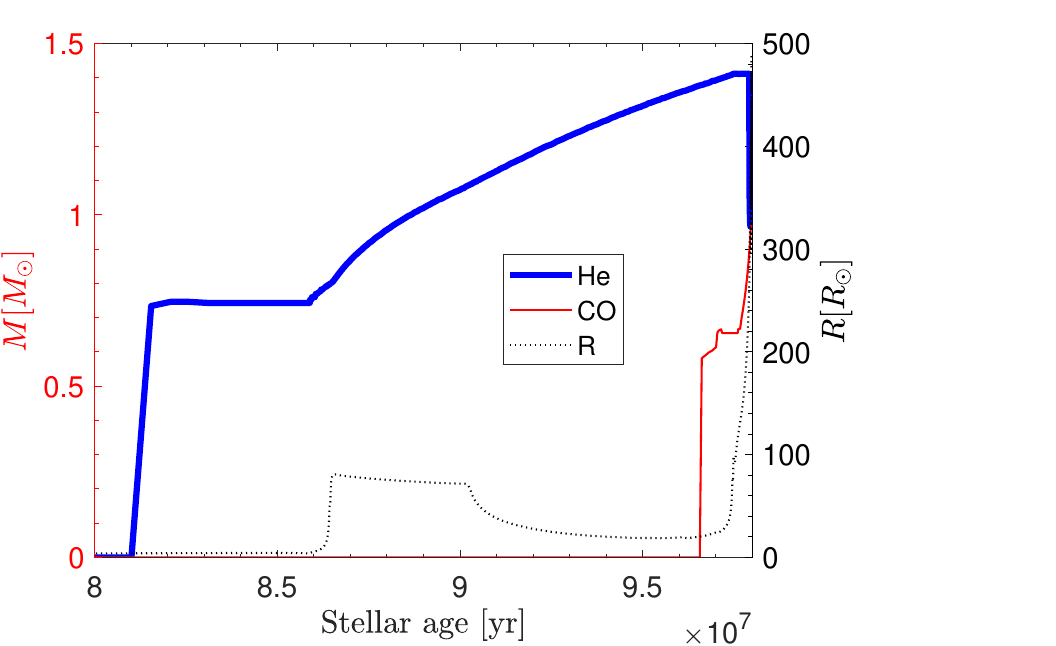} 
 \\ 
\includegraphics[trim=3.7cm 8.5cm 0.0cm 8.0cm ,clip, scale=0.59] {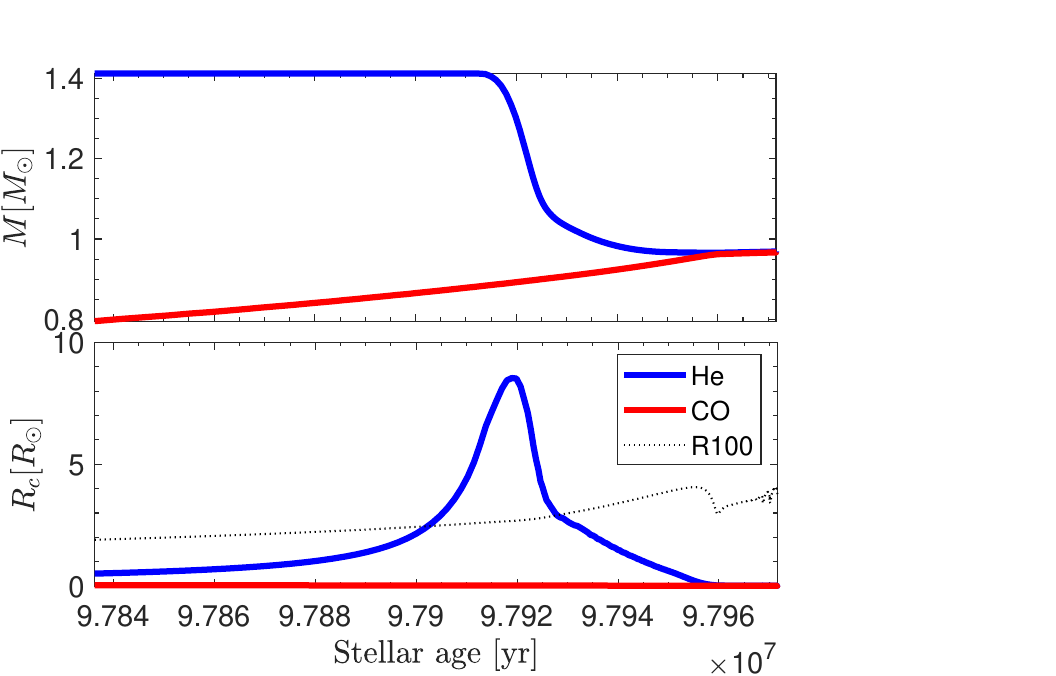}
\caption{Evolution with time at late evolutionary phases of some stellar properties for the model with a ZAMS mass of $M_{\rm ZAMS}=5 M_\odot$. The upper panel presents the mass coordinate of the CO core $M_{\rm CO}$, the mass coordinate of the He-rich core $M_{\rm He}$, and the stellar radius (dashed-black line with scale on the right).  The middle and lower panels zoom on the interesting phase when the helium-rich layer is depleted by mixing to the hydrogen-rich envelope and by nuclear burning. The dotted-black line in the lower panel is the stellar radius in units of $100 R_\odot$. 
}
\label{fig:Evolution5}
\end{figure}

\begin{figure}[]
	\centering
\includegraphics[trim=3.7cm 8.5cm 0.0cm 8.5cm ,clip, scale=0.59]{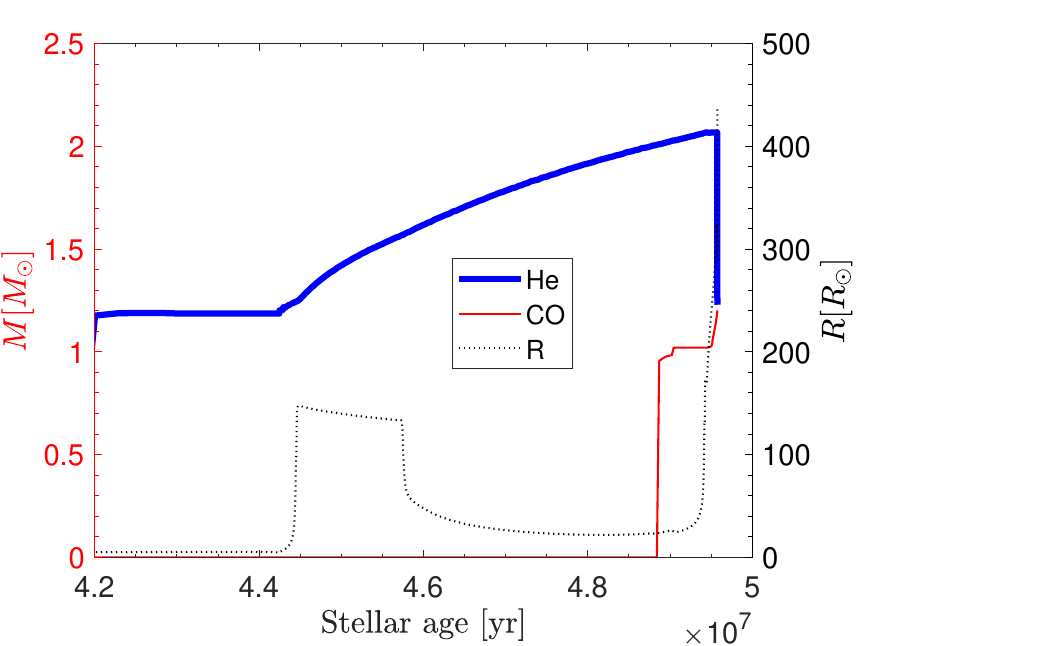}
 \\ 
\includegraphics[trim=3.7cm 8.5cm 0.0cm 8.0cm ,clip, scale=0.59]{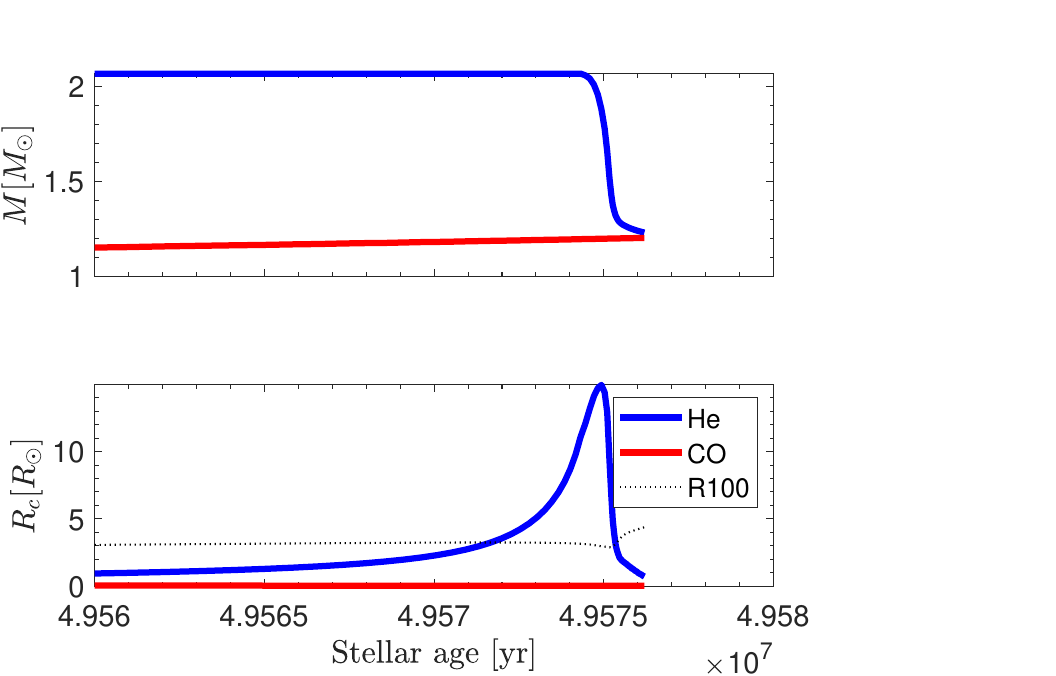} 
\caption{Similar to Fig. \ref{fig:Evolution5} but for the model with $M_{\rm ZAMS}=7 M_\odot$. }
\label{fig:Evolution7}
\end{figure}

In Figs. \ref{fig:Evolution5} and \ref{fig:Evolution7} we present the evolution with time of some stellar properties for the two stellar models, respectively, at late times that are relevant to this study. In the upper panel of each figure we present the mass coordinate of the CO core $M_{\rm CO}$, which is also the mass of the CO-rich core because we do not reach carbon ignition. We also present in the upper panel of each figure the mass coordinate of the He-rich core $M_{\rm He}$, and the stellar radius (dashed-black line with scale on the right). The mass of the helium-rich layer in the core is $M_{\rm He-r}=M_{\rm He}-M_{\rm CO}$. In the lower two panels of Figs. \ref{fig:Evolution5} and \ref{fig:Evolution7} we zoom on the late phase when the thick He-rich layer disappears because of mixing with the envelope and further nuclear burning. The middle panel shows the mass coordinates of CO and He, while the lower panel presents the radii of the CO core, of the He core (outer radius of the He-rich layer), and the stellar radius in units of $100 R_\odot$. 

 The first CEE takes place as the star rapidly expands the second time to radii of $R \ga 200 R_\odot$, while the second CEE takes place when the helium core expands. The $M_{\rm ZAMS}=5 M_\odot$ reach a radius of $R=200 R_\odot$ at $t=9.785 \times10^7 \yr$. The rapid expansion of the helium core is at $t=9.792 \times 10^7 \yr$. Therefore the time between the 
as first and second CEE is $\Delta t_{\rm 2,CEE} \la 7 \times 10^4 \yr$ (can be much shorter). For the $M_{\rm ZAMS}=7 M_\odot$ this time is $\Delta t_{\rm 2,CEE} \la 10^5 \yr$. In section \ref{subsec:CoreStripped} we find that the likely range of this timescale is $10^3 \yr \la \Delta t_{\rm 2,CEE} \la {\rm few} \times 10^4 \yr$.  

Figs. \ref{fig:HRdiagram} -  \ref{fig:Evolution7} show that for both models there is an evolutionary phase when the core obeys our requirements of $M_{\rm CO} > 0.3 M_\odot$ and $M_{\rm He-r} > 0.4 M_\odot$. The scenario that we propose in section \ref{subsec:HeRichScenario} can in principle take place if CEE takes place during this phase. Namely, a CO WD (or a HeCO WD) companion enters a CEE and removes the hydrogen-rich envelope. Thousands of years or more later it enters the He core and removes the He-rich gas to form a He-rich CSM of $M_{\rm CSM} \ga 0.3 M_\odot$. This second CEE phase might take place when the He-rich core substantially expands, up to a radius of $R_{\rm He-r} \ga 10 R_\odot$, as the lower panels of Figs. \ref{fig:Evolution5} and \ref{fig:Evolution7} show, and as we further explore in section \ref{subsec:CoreStripped}. The merger of the CO WD with the CO core leads to the formation of a WD remnant that explodes with a merger to explosion delay (MED) time of several years, allowing the He-rich wind to move away from the explosion site to a distance of hundreds of AU. To better catch the expansion of the helium-rich layer we turn to simulate the evolution after we strip the giant star from its hydrogen-rich envelope.  

\subsection{Core evolution in stripped-envelope stars}
\label{subsec:CoreStripped}

Since the qualitative behavior of the core of the $M_{\rm ZAMS} =5 M_\odot$ model (Fig. \ref{fig:Evolution5}) is similar to that of the $M_{\rm ZAMS} = 7 M_\odot$ model (fig. \ref{fig:Evolution7}), we study the evolution of the stripped envelope only for the $M_{\rm ZAMS} = 7 M_\odot$ model. 

To mimic the removal of the hydrogen-rich envelope during a CEE, the first CEE in the present scenario, we conduct a simulation for a $M_{\rm ZAMS}=7 M_\odot$ model where at $t=4.955 \times 10^7 \yr$, just left to the boundary of Fig. \ref{fig:Evolution7}, when $M_{\rm CO}=1.126 M_\odot$ and $M_{\rm He}=2.07 M_\odot$ (so $M_{\rm He-r}=2.07-1.13= 0.94 M_\odot$), we remove the hydrogen-rich envelope. We remove the mass with a constant rate of $\dot M_{\rm giant,1}= - 4.65 \times 10^{-2} M_\odot \yr^{-1}$ for $100 \yr$, and then at a lower rate of $\dot M_{\rm giant,2}= - 0.3 \times 10^{-2} M_\odot \yr^{-1}$ for a time period of $60 \yr$ in order to avoid convergence issues. 
 The leftover hydrogen mass is $M_{\rm H,S}=0.0072M_\odot$ and the total stellar mass is $M=2.093 M_\odot$. We then let the stripped core to evolve.  

\begin{figure}[]
	\centering
\includegraphics[trim=1.2cm 6.0cm 0.0cm 6.0cm ,clip, scale=0.48]{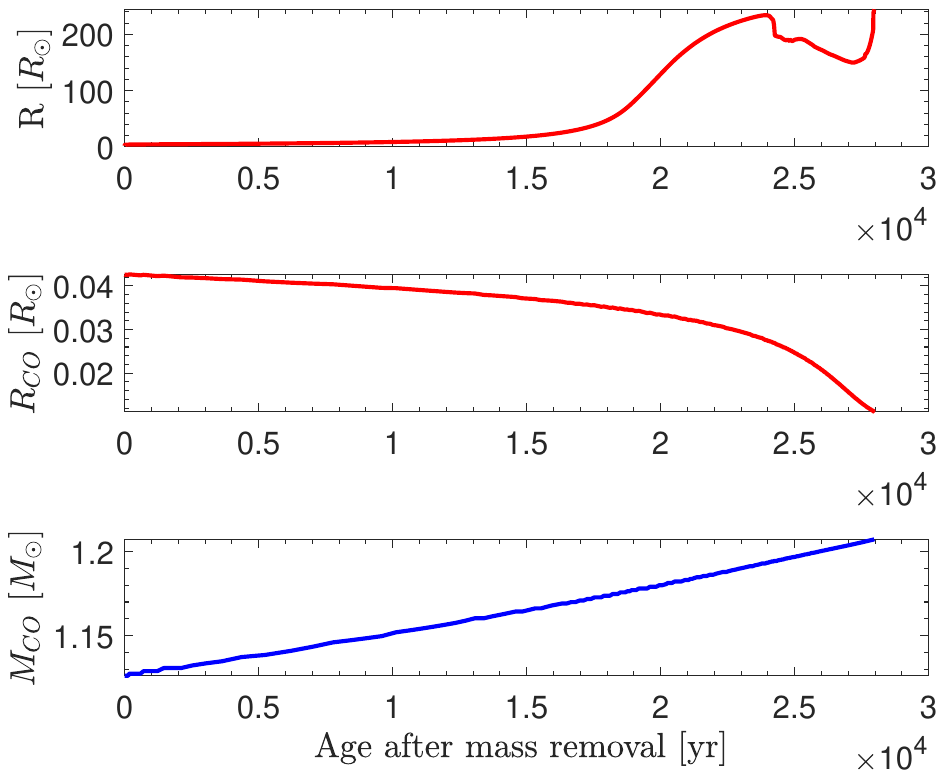}
\caption{Stellar radius (upper panel), radius of the CO core (middle panel), and the mass of the CO core (lower panel) after we removed the hydrogen-rich envelope of the $M_{\rm ZAMS} = 7 M_\odot$ stellar model at $t=4.955 \times 10^7 \yr$, leaving only $M_{\rm H,S} \approx 7.2 \times 10^{-3}M_\odot$ of hydrogen. In this graph we set $t=0$ at the end of mass removal, when the evolutionary time is  $4.9550160 \times 10^7  \yr$. }  
\label{fig:Evolution7Stripped}
\end{figure}

\begin{figure}[]
	\centering
\includegraphics[trim=3.5cm 8.5cm 0.0cm 8.0cm ,clip, scale=0.55]{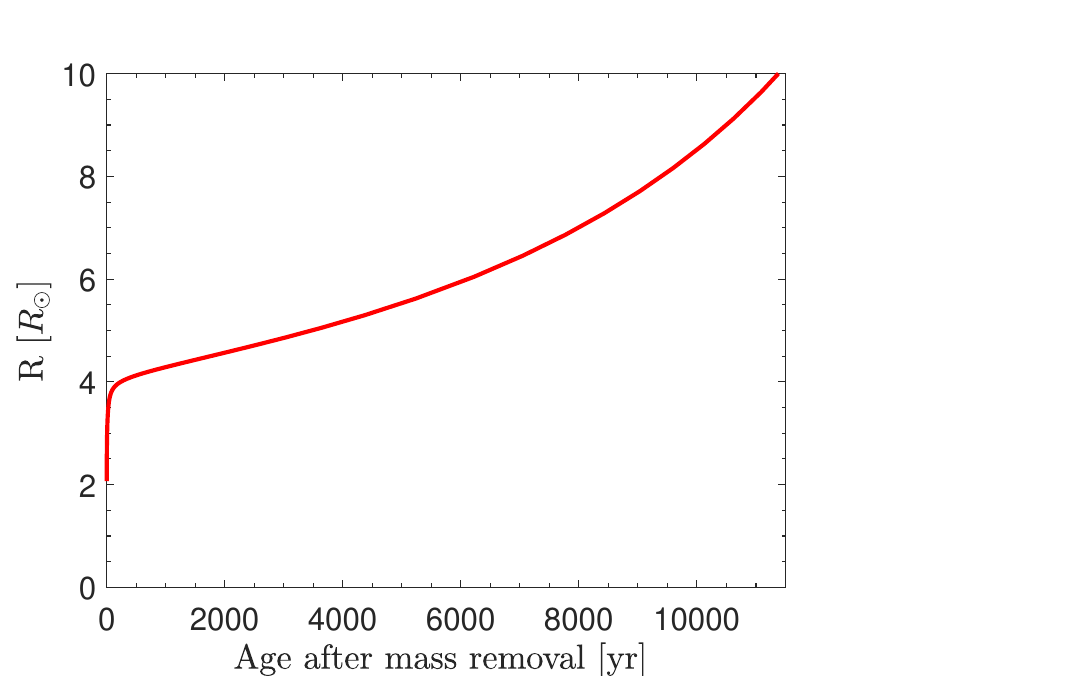}
\caption{Zooming on the early time of the upper panel of Fig. \ref{fig:Evolution7Stripped}. }
\label{fig:Evolution7StrippedZoom}
\end{figure}

In Fig. \ref{fig:Evolution7Stripped} we present the evolution of the stellar radius, and the radius and mass of the CO core, setting $t=0$ at the end of the mass removal. In Fig., \ref{fig:Evolution7StrippedZoom} we zoom on the early expansion of the star after mass removal, i.e., zooming on the upper panel of Fig. \ref{fig:Evolution7Stripped}. 
As a response to mass removal the star contracts to $R \simeq 2 R_\odot$, but immediately expands to $R\simeq 4 R_\odot$. Thereafter it is slowly expanding, e.g., reaching $R=10R_\odot$ at $t=11,527 \yr$ after mass removal.  As the CO core mass further increases and its radius decreases, the star suffers a large expansion to $R > 200 R_\odot$ at $t \simeq 2 \times 10^4 \yr$ after mass removal. However, we expect the star to engulf the WD companion at a much earlier time. 

\begin{figure}[]
	\centering
\includegraphics[trim=3.8cm 8.5cm 0.0cm 8.5cm ,clip, scale=0.58]{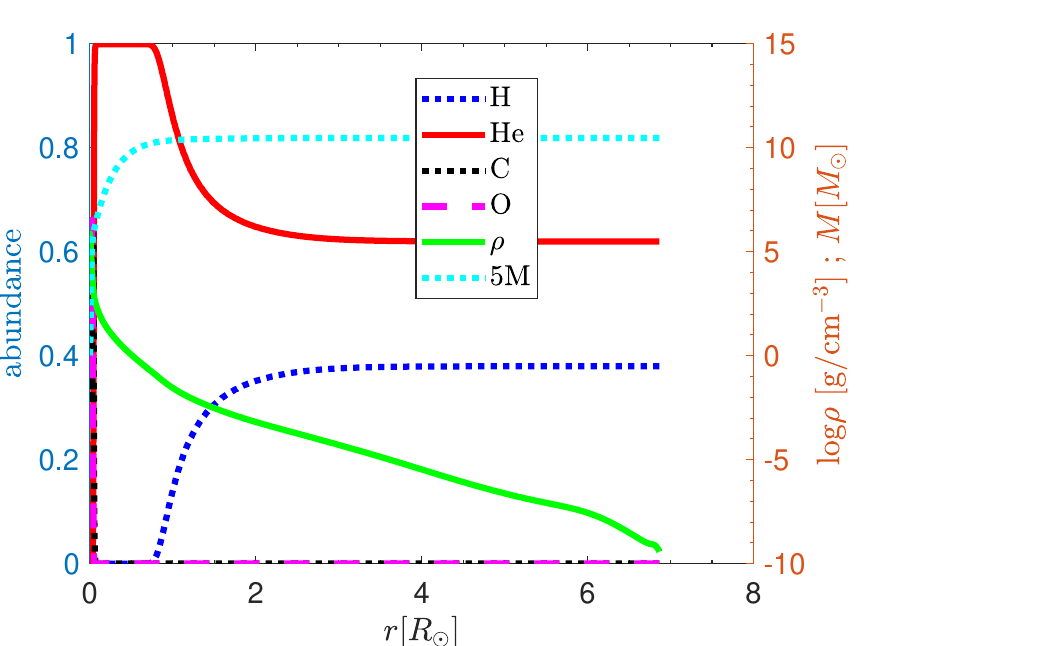}
\caption{Profiles of the mass (units of $0.2M_\odot$ on the right axis), density (logarithmic scale in units of $\g \cm^{-1}$ on the right axis), abundances (scale on the left axis), as function of radius at $t=7775 \yr$ after mass removal (mass removal was done at an age of $4.9550160 \times 10^7 \yr$). Note that the mass is in units of $0.2 M_\odot=M_\odot/5$ and therefore we mark it as $5M$, and that the total stellar mass at this time is $2.09 M_\odot$.
}
\label{fig:Profiles}
\end{figure}

In the upper panel of Fig. \ref{fig:Profiles} we present the profiles of the stellar mass (multiply by 5 to fit the scale of the graph; termed $5M$), of the stellar density, and of the relevant mass abundances as function of radius at $t=7775 \yr$ after the rapid mass removal that mimics the first CEE. It is just a representative of a time when the second CEE might start.  

Figs. \ref{fig:Evolution7Stripped} - \ref{fig:Profiles} better demonstrate our proposed CD scenario for He-rich SN Ia-CSM. A CO WD (or HeCO WD) spirals-in inside the hydrogen-rich envelope of an AGB star and removes it. It ends at $a_{\rm f1} \simeq 5-10 R_\odot$ from the core of the AGB star. 
 Spiralling-in stops or slows-down substantially because the envelope contracts to a radius of $R \simeq 4 R_\odot < a_{\rm f1}$. 
After a time period of $\Delta t_{\rm 2,CEE} \simeq 10^3 - 10^4 \yr$ the helium-rich core expands and the star engulfs the WD for the second time. The lower value of $\Delta t_{\rm 2,CEE} \simeq 10^3 \yr$ is for $a_{\rm f1} \simeq 5 R_\odot$ while the upper value of $\Delta t_{\rm 2,CEE} \simeq 10^4 \yr$ is for $a_{\rm f1} \simeq 10 R_\odot$.  Other stellar models (initial masses) and different assumptions on mass loss might give a somewhat larger time of up to $\Delta t_{\rm 2,CEE} \simeq {\rm few} \times 10^4 \yr$. But this is of no significance to the present study. 

 The expansion of the core triggers a second CEE. The WD cannot accrete much mass because the ignition of nuclear burning. Therefore, the expanding envelope engulfs the WD and due to dynamical drag it spirals-in. Future binary simulations should explore the exact nature of the second CEE phase. Here we only note that the binding energy of the envelope, from where the helium abundance is $0.9$ to the surface, when the primary radius is $R_1=6.86 R_\odot$ as the time of Fig. \ref{fig:Profiles} is $E_{\rm bind} = 1.46 \times 10^{49} \erg$. At that time the CO core mass is $M_{\rm core}=1.146 M_\odot$. For a CEE efficiency parameter of $\alpha_{\rm CEE}$ the spiralling-in WD ends the secon d CEE at a radius of
\begin{equation}
a_{\rm f2}= 0.027\left(\frac{\alpha_{\rm CEE}}{0.3} \right)
\left( \frac{M_{\rm WD}}{0.6 M_\odot} \right) R_\odot.
 \label{eq:jacc}
\end{equation}
Making the calculation for when the radius of the primary is anywhere from $R_1=4.5 R_\odot$ to $R_1=10 R_\odot$ does not change the value of $a_{\rm f2}$ by more than $10\%$. 
Since the radius of the CO core at that time is $R_{\rm core}=0.04 R_\odot$, we conclude that the second CEE might lead to a WD-core merger, depending on the value of $\alpha_{\rm CEE}$.

By the merger time the hydrogen-rich ejecta is at a large distance of $\approx 1 \pc$. After a merger to explosion delay (MED) time the core explodes, as the CD scenario assumes.     
For the specific case of SN 2020eyj, where the inner boundary of the He-rich CSM is at $r_{\rm CSM,in}\simeq 4 \times 10^{15} \cm$ \citep{Kooletal2022}, the MED time is $t_{\rm MED} \approx 10 \yr$ for a helium expansion velocity of $\approx 100 \km \s^{-1}$.  
    
\section{Summary}
\label{sec:Summary}

The motivation of our study is the new observations of SN 2020eyj, a SN Ia-CSM with a helium-rich CSM \citep{Kooletal2022} and the need to consider all SN Ia scenarios when analysing observations, as the long list of recent papers that study different scenarios suggests (section \ref{sec:intro}). 
\cite{Kooletal2022} dismiss the DD scenario for SN 2020eyj and argue for the SD scenario. However, they did not consider the CD scenario. In section \ref{sec:SN2020eyj} we argued that the SD scenario has some difficulties to account for some properties of SN 2020eyj, which beside the presence of a compact CSM is a normal, on the faint side, SN Ia. 

Building on earlier papers that argue for the CD scenario for SNe Ia-CSM (section \ref{sec:SN2020eyj}) we propose the CD scenario also for SN 2020eyj, but we consider a new channel that accounts for the helium-rich CSM (section \ref{sec:TheCDscenario}). 

While in the channel of the CD scenario for hydrogen-rich SN Ia-CSM the CEE ends with the merger of the WD companion with the core shortly after the onset of the CEE, in the new channel there are two major CEE phases (section \ref{subsec:HeRichScenario}). After the WD removes the hydrogen-rich envelope of the AGB star the spiralling-in process ends  at an orbital separation of $a_{\rm f1} \simeq 5-10 R_\odot$.  Only hundreds to about $10^4 \yr$ years later, after further tidal interaction and, mainly, after the helium-rich core expands (Figs. \ref{fig:Evolution5} - \ref{fig:Evolution7StrippedZoom}) the WD enters a CEE with the helium-rich layer. By that time the hydrogen-rich envelope is at $\approx 0.1- 1 \pc$ and therefore the ejecta collides with a helium-rich CSM at hundreds of AU. The WD merges with the core during the second CEE phase, and only after a time (the MED time) of weeks to tens of years the merger remnant explodes. 
 
In section \ref{subsec:Core} we followed the evolution of two stellar models (Figs. \ref{fig:Evolution5} and \ref{fig:Evolution7}). The evolution shows that during the second expansion of the primary star and while it is still hydrogen-rich, i.e., towards its final AGB, there is a massive CO core, there is a massive helium-rich layer, and the core is compact. This allows the first CEE to remove the hydrogen-rich envelope but for the WD companion to stay outside the core. We further demonstrated this in section \ref{subsec:CoreStripped} by removing the hydrogen-rich envelope and letting the stripped core to evolve. The result is an expansion of the massive helium-rich layer (Figs. \ref{fig:Evolution7Stripped} and \ref{fig:Evolution7StrippedZoom}). This expansion can force the WD to spiral-in and merge with the CO core. According to the CD scenario the merger product might explode later as a SN Ia. 
  
There are two important ingredients in the helium-rich SN Ia-CSM CD scenario. The first is that the WD companion enters a CEE with the AGB star during a phase of a massive CO core and a massive helium-rich compact layer such that the WD can spiral-in and remove the hydrogen-rich envelope before merging with the core, but it does not end at a too large orbital separation as to prevent the core-WD merger during the second CEE phase. This hydrogen-rich envelope removal allows a time gap of hundreds to tens of thousands of year between the first and second CEE phases. The second ingredient is that after the WD merges with the CO core and the second CEE phase ends, the MED time, i.e., the time from merger to explosion, is weeks to tens of year such that the ejecta collides with the helium-rich CSM within days to months after explosion. In the case of SN 202eyj we estimate the MED time to be $\approx 10 \yr$.  

In this study we presented the feasibility of the helium-rich SN Ia-CSM CD scenario. In a future study we will follow the entire binary evolution, e.g., with \textsc{mesa}-binary, to examine whether a core-WD merger can indeed take place after a second CEE phase.

\section*{Acknowledgments}

 We thanks an anonymous referee for very useful comments, in particular on the mass removal simulation.   This research was supported by a grant from the Israel Science Foundation (769/20).

\section*{Data availability}
The data underlying this article will be shared on reasonable request to the corresponding author.  


\end{document}